\documentstyle[12pt,epsf]{article}
\def\fun#1#2{\lower3.6pt\vbox{\baselineskip0pt\lineskip.9pt
  \ialign{$\mathsurround=0pt#1\hfil##\hfil$\crcr#2\crcr\sim\crcr}}}
\relax
\makeatletter
\@addtoreset{equation}{section}
\makeatother

\let\k=\kappa
\let\m=\mu\let\n=\nu

\let\z=\zeta\let\G=\Gamma
\let\D=\Delta

\newcommand{\be}{\begin{equation}}
\newcommand{\ee}{\end{equation}}
\newcommand{\bea}{\begin{eqnarray}}
\newcommand{\eea}{\end{eqnarray}}

\newcommand{\nbox}{{\,\lower0.9pt\vbox{\hrule \hbox{\vrule height 0.2 cm \hskip
0.2 cm \vrule height 0.2 cm}\hrule}\,}}

\advance\textheight by \topskip
\begin{document}
\begin{titlepage}

\vspace{-1.5truecm}

\rightline{SCIPP97/28}
\rightline{SLAC-PUB-7676}
\rightline{\tt hep-th/9710174}

\begin{center}
{\Large\bf
Multigraviton Scattering in the Matrix Model \\
}
\vskip 0.5 cm
{\large {\bf
M. Dine${}^{a}$
\footnote{\tt
dine@scipp.ucsc.edu
}
,
A. Rajaraman${}^b$
\footnote{\tt
arvindra@leland.stanford.edu
}
\footnote{Work supported in part by the Department of Energy
under contract no. DE-AC03-76SF00515.
}
}}\\
\vskip 0.75 cm
{\it ${}^a$Santa Cruz Institute for Particle Physics, University of California,
Santa Cruz, CA~~95064\\
 ${}^b$Stanford Linear Accelerator Center, Stanford University,
Stanford, CA 94309\\
}

\end{center}
\begin{abstract}
We consider scattering processes in the matrix model
with three incoming and three outgoing gravitons.
We find a discrepancy between the amplitude calculated from
the matrix model and the supergravity prediction.
Possible sources
for this discrepancy are discussed.
\end{abstract}
\end{titlepage}

\section{Introduction}

One of the exciting developments to emerge
from the work in recent years on string duality is the
conjecture that in the infinite momentum frame,
string theory (M theory) is described by a large $N$
matrix model \cite{bfss}\ . In addition, the finite $N$ matrix
model has been conjectured to describe M-theory on a
compact light-like circle \cite{DLCQ}\ .

 In eleven dimensions, this matrix model
is just a supersymmetric quantum mechanics with $16$
supercharges. In dimensions $8-10$, it is a field theory \cite{susskindtorus,
taylortorus, shrink}.
In lower dimensions, the story is more complicated \cite{roz,
berkroz, seib1}.

The matrix model conjecture
has been well tested in
processes involving scattering of two gravitons \cite{joeb} and
other two-body interactions \cite{lifsch, tseyt, joea}.
Matrix models also reproduce known string dualities
in remarkable ways, and seem able to reproduce the
light-cone three string vertex \cite{dvv}.  Seiberg
has recently explained how the conjecture might be
derived \cite{nati}.

In this note, we describe a further test. 
We will compute, in this note, an amplitude
in
supergravity involving six gravitons -- three incoming,
three outgoing -- and compare this to a calculation in
M(atrix) theory. We will work in the limit
of small velocities and very small momentum transfer,
with zero longitudinal momentum exchange.  This is analoguous to the
well-known computation of the $v^4/r^7$ force law
between two gravitons as a one-loop computation
in the matrix model \cite{bfss, dkps, bachas}.  A rough estimate suggests
that the matrix model is on the right track.  If all three
gravitons are separated by a similar distance,
$R$, then the $3 \rightarrow 3$ amplitude should
behave as $1 \over R^{14}$, and this is indeed
the behavior one obtains by simple power counting
on the matrix model diagrams.

However, when the amplitude
is examined in more detail, we seem to find
a difficulty.
We will take a limit where one graviton is far from the other
two, i.e. $ r_{13}\sim r_{23}=R \gg r_{12}=r$.  We
will expand the amplitude in powers of  $1/R$.
In momentum space this corresponds to a limit in which
one of the momentum transfers, say $q_1$, is much
smaller than the other two ($q_2$, $q_3$).

Since the propagator for a graviton goes as
$1/R^7$ (there is no longitudinal momentum
transfer), we expect the leading behaviour to go
as ${1\over R^7r^7}$. The vertices are
bilinear in momenta, so we might expect a term in
the amplitude of the form
 \be {(k_1\cdot k_2)(k_2\cdot k_3)(k_3\cdot k_1)
\over q_1^2q_2^2}
\ee
where $k_1,k_2,k_3$ are the momenta of the incoming
gravitons.

In the next section we argue by simple power counting
on the appropriate Feynman diagrams
that such a term can not
occur in the matrix model.
If the matrix model is truly to reproduce
the supergravity amplitude,
then
the coefficient of the term in the supergravity amplitude
must be zero.

The third and fourth sections are devoted to calculating the
coefficient of the term in supergravity . We find that the
coefficient is nonzero. We first do the calculation using
the Feynman diagrams in supergravity; as a check, we then
perform the calculation in string theory. The two computations
agree.

Finally, we speculate on possible origins of the discrepancy we have
found.

\section{The matrix argument}.

In the matrix model, it is easiest to study scattering
with zero $p^+$ exchange,  and at large
impact parameter, corresponding to large expectation
values for the $\vec x$'s.  In momentum space,
this means one studies momentum transfers small
compared to the momenta themselves.  If the incoming
and outgoing momenta of the first graviton are $k_1$ and
$k_2$, respectively, then $k_1 = k_2+q_1$,
where $\vert \vec q_1\vert  \ll \vert \vec k_1\vert$, $q^+=0$.
Similar remarks apply to the momenta $k_3,k_4$
and $k_5,k_6$ associated with the other lines.

Among the invariants relevant to this problem are the
invariant energies associated with the various subsystems,
$k_1 \cdot k_3$, $k_1 \cdot k_5$, $k_3 \cdot k_5$.
In light cone variables, these have a simple form.  For example,
\be
k_1 \cdot
k_3 = k^+_{1}k^+_{3} (v_1-v_3)^2.
\ee

The terms we are looking for in the amplitude, then
have the form
\be
 {(k_1 \cdot k_2)(k_2\cdot k_3)(k_3 \cdot k_1)
\over R^7r^7} =(k^{+2}_{1} k^{+2}_{2}
k^{+2}_{3} ) {(v_1-v_2)^2(v_1-v_3)^2(v_3-v_2)^2
\over R^7r^7}.
\ee

We want to ask how such a term can arise in the matrix model.
Without loss of generality, take $\vec x_3=0$, $\vert x_2\vert=r$, $\vert x_1
\vert
 = R$.
In order to generate a term in the matrix model of the form above,
we must study
loop graphs with the following properties:
\begin{itemize}
\item They must depend on two masses ( $R$ and $r$) so they must contain
at least two
loops.
\item They must have six external leg insertions;
two for $(v_1-v_2)$, two for $(v_1-v_3)$,
and two for $(v_3-v_2)$.
\item
The four factors of $v_1$ all attach to ``heavy'' modes,
with mass (frequency) of order $R$.
\end{itemize}
An example of such a graph is shown in Fig. 1.
\begin{figure}
\centerline{\epsfysize=5cm \epsfbox{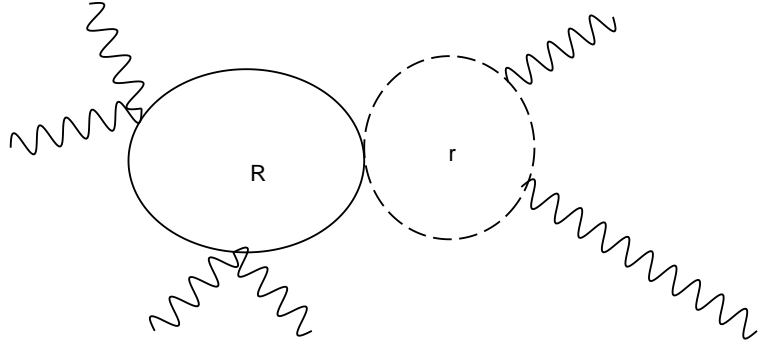}}
\caption{One of the Feynman diagrams of the matrix model.}
\end{figure}

Simple power counting shows that this graph (and all of the others
with these properties) are suppressed
by at least nine powers of $R$.
The three propagators each give $1/R^2$; extracting
four velocities gives $1/R^4$, and the loop integration
gives  one power of $R$. Thus the graph goes
as $1/R^9$.  Supersymmetry cancellations might give further
suppression, but already this is smaller than $1/R^7$.
An identical suppression can be found in the background
field method used by \cite{lifsch, joeb}.

One can alternatively do the counting in terms of effective
operators.  Integrating out first the more massive fields,
i.e. integrating over the $R$ loop, yields an effective
local operator built of the light fields.  Since $x_1$
doesn't couple directly to light fields, this operator
must be of the form
$(v_1-v_2)^2(v_1-v_3)^2 X^2$. Here $X$ represents
operators which can couple to the light fields.
On dimensional grounds, this should go as $1/R^9$.
The integral over the light field loop cannot induce
compensating powers of $R$.

We can also ask what sorts of terms are generated by
the iteration of the one loop matrix model Hamiltonian.  It is easy
to show that there are no terms of the correct form
here as well.  We will return to this point in our concluding
remarks.  One can also easily check that higher loop contributions
are further suppressed.

In short, we see that no term of the form
\be
 {(k_1 \cdot
k_2)(k_2  \cdot k_3)(k_3 \cdot k_1)
\over R^7r^7} \propto  {(v_1-v_2)^2(v_1-v_3)^2(v_3-v_2)^2
\over R^7r^7}\ee
can be generated in the matrix model at any loop order.

\section{Supergravity}

We now turn to the calculation in supergravity.  In eleven dimensions,
this just means that we need to consider graviton exchange between
the external gravitons, so the required vertices can be read off the
Einstein lagrangian.
The relevant diagrams are shown in Figure 2.
\begin{figure}
$$
\begin{array}{ccc}
\hskip -5cm
\epsfxsize=40mm
\epsfbox[0 50 200 200]{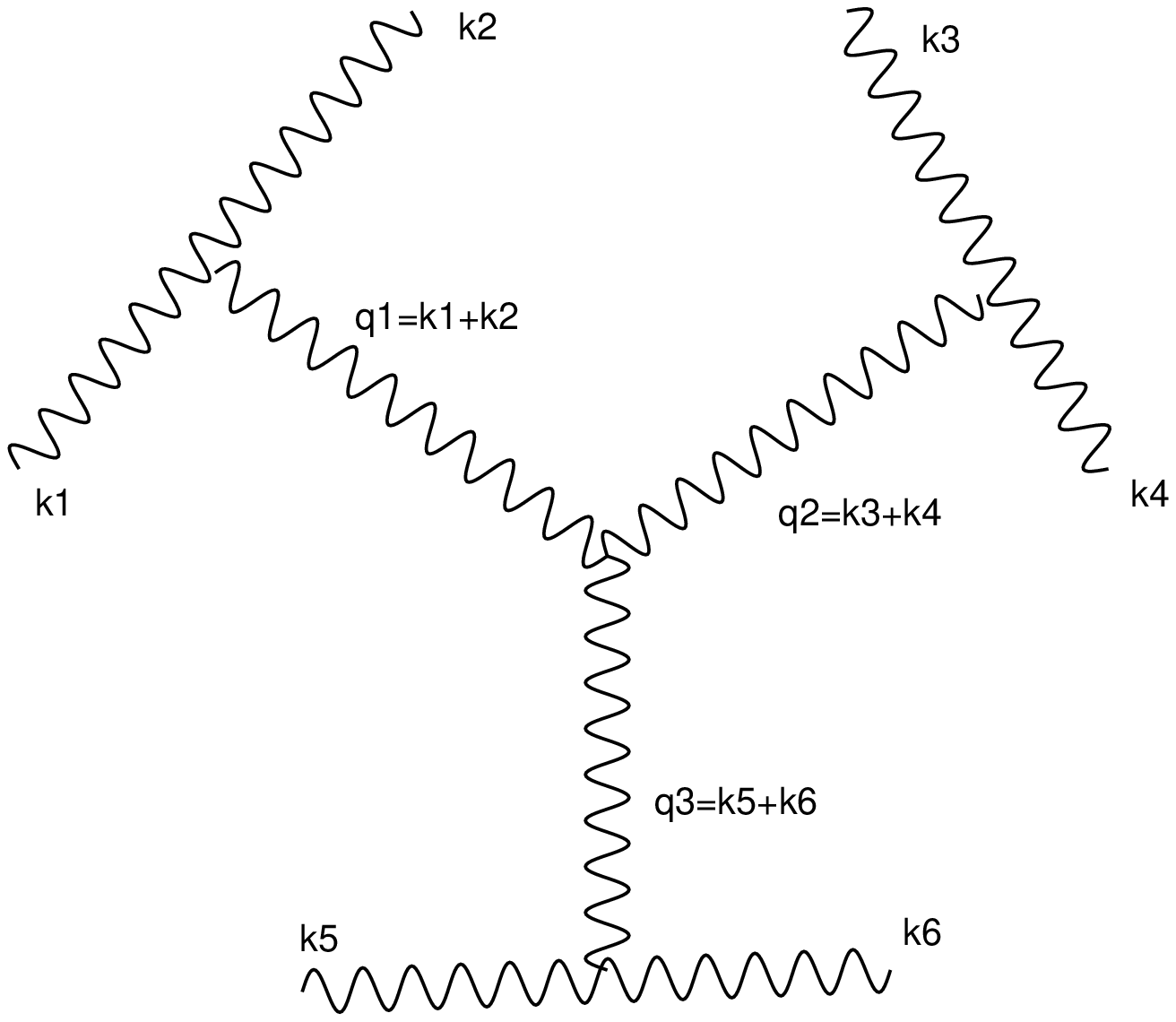} & \epsfxsize=45mm
\hskip5cm
\epsfbox[0 50 200 200]{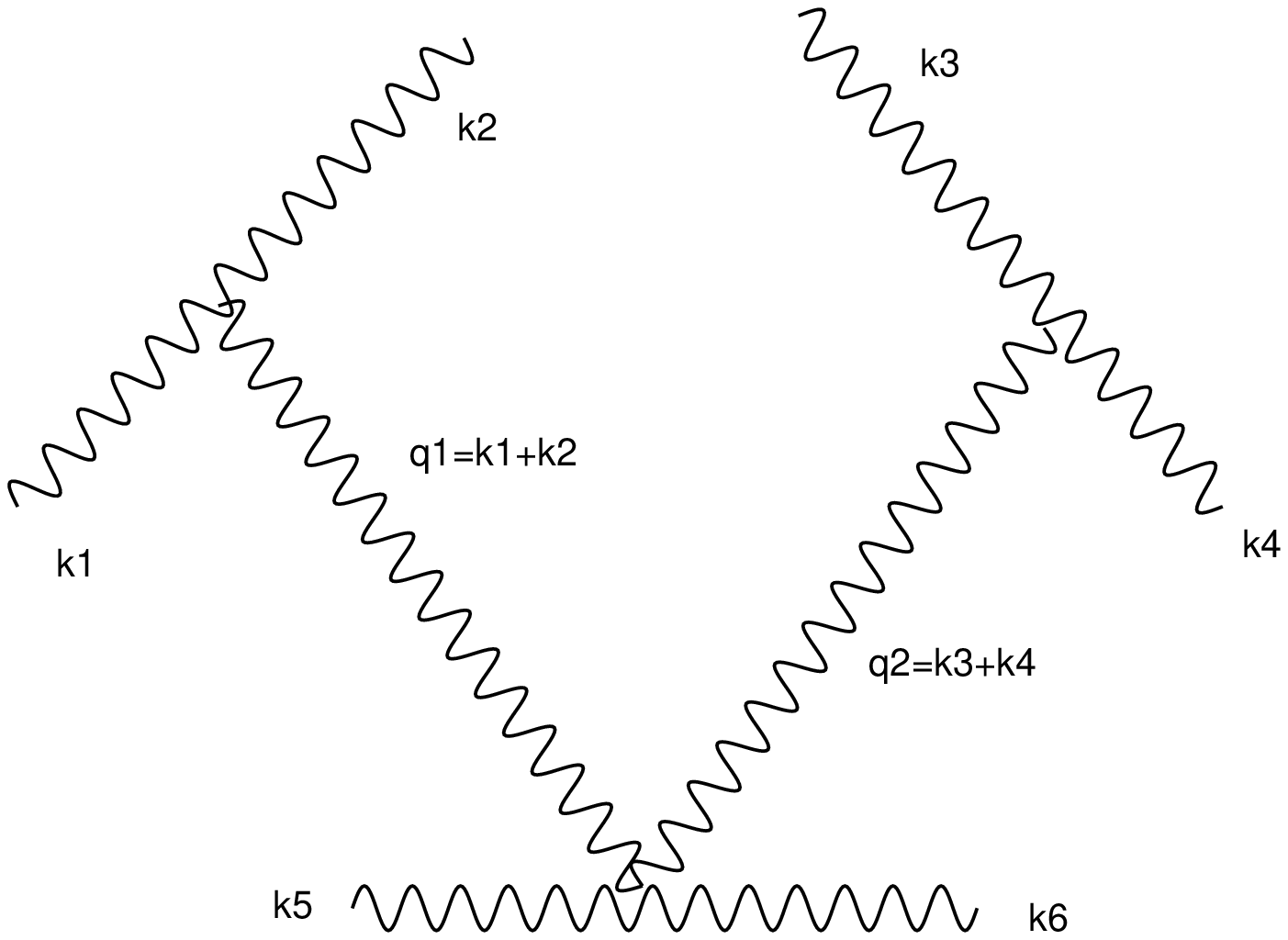}
\end{array}
$$
\medskip
\caption{The supergravity graphs }
\end{figure}
\medskip


To evaluate the graphs, we need to know the structure
of the supergravity vertices.  These are conveniently
collected in \cite{sanan}.  The
three graviton vertex appears in eqn. 2.6 of that
paper, the four graviton
vertex in 2.8.
Here we give a brief summary of the computation.

Work in momentum space.  Immediately take the limit
$q_2\approx -q_3 \gg q_1$, where $q_i$ are
the momentum transfers.  The leading terms in the matrix
model don't involve the fermionic variables and thus
don't change the polarizations; helicity-changing
terms are suppressed by powers of the impact parameter
(or the momentum transfer, in momentum
space).  So we look for terms
in the graviton amplitude
\begin{equation}
k_1 \cdot k_3 k_1 \cdot k_5 k_3 \cdot k_5 \xi_1 \cdot \xi_2
\xi_3 \cdot \xi_4\xi_5 \cdot \xi_6\over q_1^2 q_3^2 \label{eq:kinematics}
\end{equation}

The vertices
where an incoming graviton emits a single, slightly
off shell graviton are particularly simple.  Calling the
incoming and outgoing momenta $k_1$ and $k_2$, respectively,
and $q$ the momentum transfer, and taking the polarization
indices to be $\alpha \mu$, $\beta \nu$, $\sigma \gamma$,
the only relevant terms (not suppressed by powers of $q$)
are
$$P_3( k_{1 \sigma} k_{2 \gamma} \eta_{\mu \nu} \eta_{\alpha \beta})
+ P_6(k_{1 \sigma} k_{1 \gamma} \eta_{\mu \nu} \eta_{\alpha \beta}).$$
The $P_i$'s denote symmetrization in the three gravitons; the subscript
denotes the number of distinct permutations.  Note that in the
limit of interest, the $q$'s are small compared to the transverse
momenta, so we need to keep in every instance the smallest possible
number of $q$'s.
In the first term, then, the only relevant symmetrization is $1\rightarrow 2$,
but this gives nothing new; in the second, this does.  In
this limit, $k_1 \approx -k_2$ (all momenta are defined as
flowing into the vertex) so we have simply
\begin{equation}
V_a \approx k_{1 \sigma} k_{1 \gamma} \eta_{\mu \nu} \eta_{\alpha \beta}.
\end{equation}

Next consider the vertex involving three off shell gravitons.
For the kinematic structure of equation
(\ref{eq:kinematics}),
We are only interested in the term involving $q_2 \cdot q_3 \approx
-q_3^2$.
There is only one such term:
$$-2 P_3(q_1 \cdot q_3 \eta_{\alpha \nu} \eta_{\beta \sigma}
\eta_{\gamma \mu})$$
where the gravitons $1,2,3$ carry indices $\mu \alpha$,
$\nu \beta$, $\sigma \gamma$, respectively.
The symmetrization produces no additional factors.

Thus for the diagram with the three graviton vertex we obtain
\begin{equation}
(i\k^4)(\xi \cdot \xi)^3 (k_1\cdot k_3)(k_1\cdot k_5)(k_5\cdot k_3)\left( {1\over
q_1^2q_3^2}\right)
\end{equation}

The second diagram is the only other relevant one for this
structure.  We need the four graviton vertex appearing here.
For such a vertex, label the incoming lines
$k_1,\mu,\alpha;k_2 \nu,\beta;
q_3,\sigma,\gamma;q_4, \rho, \lambda$.  Because this diagram
already has $1 \over q_3^2   q_1^2$, we can ignore all terms
of order $q$ at the vertex.  In addition, we must again insure that the
helicity on the incoming and outgoing graviton lines is unchanged.  A careful
examination of the
expression for the four graviton vertex indicates that there
are only two relevant terms,
$$-P_{12}(k_{1 \sigma} k_{2\rho} \eta_{\gamma \lambda}
\eta_{\mu \nu} \eta_{\alpha,\beta}) -2P_{12}(k_{1,\nu}
k_{1,\sigma} \eta_{\beta \gamma} \eta_{\mu \rho} \eta_{\alpha\lambda}).$$
The second term does not obviously preserve helicities.  However, if one
exchanges $2 \rightarrow 4$, then one obtains
$$-P_{12}(k_{1 \sigma} k_{2\rho} \eta_{\gamma \lambda}
\eta_{\mu \nu} \eta_{\alpha,\beta}) -2P_{12}(k_{1,\rho}
k_{1,\sigma} \eta_{\lambda \gamma} \eta_{\mu \nu} \eta_{\alpha\beta}).$$
In the first term, exchanging $1 \rightarrow 2$ gives an inequivalent
result (but exchanging $3 \rightarrow 4$ does not). In the second term,
exchanging $1$ and $2$
similarly gives an inequivalent result.  So at the vertex
we have, using $k_1=-k_2$,
$$-2 k_{1,\rho}
k_{1,\sigma} \eta_{\lambda \gamma} \eta_{\mu \nu} \eta_{\alpha\beta}.$$
One can now combine this with the three graviton vertices to obtain for the
diagrams (note that in this limit, the four-graviton vertex can attach to
either
of two of the external lines, giving the same result)
\bea
(-2i\k^4)(\xi \cdot \xi)^3 (k_1\cdot k_3)(k_1\cdot k_5)(k_5\cdot k_3)\left( {1\over
q_1^2q_3^2}\right)
\eea

The two contributions do not cancel.

We have done many further checks on these amplitudes.  In the
next section, we will show that this amplitude agrees with a
string computation.  It is easy to compute many of the other
terms in the vertex and to compare these with a string computation
as well, and we have done several additional tests of this sort.

\section{Strings}

We now turn to the string calculation. We will work in bosonic string theory.
The M-point tensor amplitude is given by \cite{kawai}
\bea
A_{\m_1\n_1..\m_M\n_M}\z_1^{\m_1\n_1}..\z_M^{\m_M\n_M}&&=\pi\k^{M-2}
\int d^2z_1..d^2z_m {1\over \D}
\nonumber\\
&&\prod_{i>j}{(z_i-z_j)^{k_{ij}}}
exp\left[ \sum_{i>j}{{\z_i.\z_j\over (z_i-z_j)^2}}-\sum_{i>j}{{k_i.\z_j\over
2(z_i-z_j)}} \right]
\nonumber\\
&&\prod_{i>j}{(\bar{z}_i-\bar{z}_j)^{k_{ij}}}
exp\left[ \sum_{i>j}{{\bar{\z}_i.\bar{\z}_j\over
(\bar{z}_i-\bar{z_j)^2}}-\sum_{i>j}{{k_i.\bar{\z}_j\over
2(\bar{z}_i-\bar{z}_j)}} }\right]
\eea
where
\be
\D={d^2Z_ad^2Z_bd^2Z_c\over | Z_a-Z_b|^2| Z_a-Z_c|^2| Z_c-Z_b|^2}
\ee
means that three variables are to be given arbitrary values and not integrated.
Here we have introduced
\be
k_{ij}=k_i\cdot k_j/4
\ee
We are interested in terms where the polarizations dot into themselves, thus we
need
\be
\z_1.\z_2 \bar{\z}_1.\bar{\z}_2 \z_3.\z_4 \bar{\z}_3.\bar{\z}_4 \z_5.\z_6
\bar{\z}_5.\bar{\z}_6
\ee
We will in addition choose $z_6=\infty,z_5=1,z_4=0$. This means that all terms
involving $z_6$ cancel.

The resulting expression has the structure (we will omit the integration symbol
henceforth)
\bea
z_{12}^{k_{12}-2}z_{13}^{k_{13}}z_{14}^{k_{14}}z_{15}^{k_{15}}
z_{23}^{k_{23}}z_{24}^{k_{24}}z_{25}^{k_{25}}
z_{34}^{k_{34}-2}z_{35}^{k_{35}}
\eea
multiplied by its complex conjugate .

We are interested in poles involving ${1\over k_{12}}$, ${1\over k_{34}}$,
${1\over k_{56}}$.  To isolate singularities of this sort,
the following change of variables is helpful:
\bea
z_1= (r_1+1)z_2~~~~~z_2=z_2   ~~~~~~z_3=r_3 z_2
\eea
The integrand then becomes
\begin{equation}
z_2^{-2+k_{56}}r_1^{-2+k_{12}}r_3^{-2+k_{34}}f(r_1,r_3,z_2)
\label{integrand}
\ee
where

\bea
f(r_1,r_3,z_2)= (1+r_1-r_3)^{k_{13}}(1+r_1)^{k_{14}}
(1-z_2-r_1 z_2)^{k_{15}}(1-r_3)^{k_{23}}
(1-z_2)^{k_{25}}(1-r_3z_2)^{k_{35}}
\eea
In this form, the poles of interest come from the integration
regions where the variables go to zero.

Indeed, it is now a simple matter to expand the function $f$ in,
say, $r_1$,$r_3$.  Keeping the first order terms isolates
the pole $(k_{12}k_{34})^{-1}.$  The remaining $z_2$ integral
is readily done exactly.   The most singular term,
proportional to $({k_{12}k_{34}k_{56}})^{-1}$
agrees with the supergravity calculation.  The term which
interests us, proportional to
$k_{13}k_{15}k_{35}(k_{12}k_{34})^{-1}$ is non-vanishing and
also agrees with the supergravity result.
A more detailed study of
this integral shows that the properly
normalized amplitude reproduces our supergravity result.
Those attempting to verify these statements may find the integral:
\bea
&&I(n_1,n_2,n_3,n_4)=\int d^2z
z^{A-n_1}(1-z)^{B-n_2}\bar{z}^{A-n_3}(1-\bar{z})^{B-n_4}
\nonumber\\
&&=\pi (-1)^{n_4}{
\G(A+1-n_1)\G(B+1-n_2)\G(n_3+n_4-A-B-1)\G(B+1-n_4)
\over \G(-B)\G(1+B) \G(A+B-n_1-n_2+2)\G(n_3-A)}
\eea
helpful.
As noted earlier, we have also checked many of the other
terms in the amplitude against the supergravity computation.


\section{Implications}

We have not found a satisfactory explanation of the discrepancy
we have found. However, there are several potential issues
in the calculation.

We must  emphasize that we have done a matrix model calculation at a
finite value of  $N$. This means that we have to compare it to M-theory
with a compact lightlike circle as proposed in \cite{DLCQ} . One might worry
that this brings in
complications while evaluating the supergravity amplitude. However,
one expects that for a large value of the light-like radius $R$ (or,
equivalently, a large
value of $N$), the theory on a compact lightlike
circle tends to classical  11-dimensional supergravity with corrections
that are suppressed as ${1\over R}$.  Hence we
do not expect corrections from the lightlike compactifications to
alter our supergravity result (as the corrections scale differently
with  $R$ and $N$.) It is possible that this is too naive
and that DLCQ M-theory contains terms which are not
present in the classical supergravity lagrangian.  However, this seems to
lead to
the counterintuitive
conclusion  that DLCQ  M-theory does not
behave as classical supergravity in the low energy limit.

 It is also  possible that
we have blundered in some way in our evaluation of
the classical amplitude.  This would be the
simplest resolution of the puzzle.
But we have performed numerous checks of the Feynman rules, as well
as of our approach to treating the string computation.  The detailed
agreement of the two computations is impressive.

Another concern is that
the iteration of the one loop matrix model Hamiltonian
will generate contributions to the three graviton
scattering amplitude.  However, these have the wrong
$N$ and $R$ dependence to resolve the discrepancy.
Indeed, the most singular parts
of these graphs are easily seen to
reproduce the Feynman diagrams in which a graviton
is first exchanged, say, between the first and second
line, and then another graviton is exchanged between
the second and third.

Let us first review how the $N$ counting goes for the one
loop diagram, corresponding to the $2 \rightarrow 2$
process.   On the matrix model side, the result is
proportional to ${N_1 N_2 \over R^3} (v_1-v_2)^4/t$.  How does this
compare to the supergravity calculation?  The
invariant amplitude behaves as
$\kappa^2(k_1 \cdot k_2)^2/t$.  However, this is for relativistically
normalized states.  To go to non-relativistic normalization,
we need to divide by a factor of $\sqrt{k^+}$ for
each external line.  This leaves us with
$$\kappa^2 k^+_1k^+_2(\vec v_1-\vec v_2)^4
={N_1 N_2 \over R^3 M_p^9}(\vec v_1-\vec v_2)^4.$$
This is exactly as above.

\begin{figure}
\centerline{\epsfysize=5cm \epsfbox{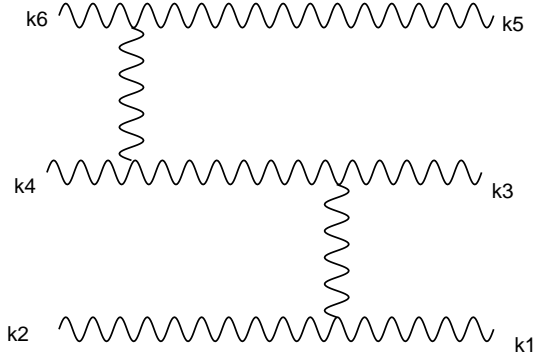}}
\caption{The gravity graph corresponding to the one-loop iteration.}
\end{figure}

Now let's consider the Feynman diagrams in the $3 \rightarrow 3$ process.
First consider the supergravity side.  The invariant
amplitude is proportional to
$(k_1 \cdot k_3)^2(k_3 \cdot k_5)^2 /k_3 \cdot q_3$,
where $q_3$ is the small momentum transfer, and we have used
the fact that $k_3^2=0$.  In terms of light cone variables,
this is
$$(k^+_{1} k^+_{3})^2
(v_1-v_2)^4 (k^+_{3}k^+_{5})^2(v_2-v_3)^4
\over k_3 \cdot q_3$$
The external state normalization factors give
$(k^+_{1}k^+_{3}k^+_{5})^{-1}$.  In terms of $N$-dependence,
this leaves
$${N_1 N_2^3 N_3
(v_1-v_2)^4 (v_2-v_3)^4
\over R^{5}k_3 \cdot q_3}$$

It is easy to see how this is reproduced by the matrix model.
The iteration of the lowest order Hamiltonian reproduces the velocity
factors and gives $N_1 N_2^2 N_3/R^6$.  The energy denominator is
$(\vec k_3 \cdot \vec q_3)/N_2R^{-1}$.  So we obtain the amplitude
above.

One could imagine all sorts of corrections to the matrix
model Hamiltonian (at one loop) which would give
things like $\vec q_3 \cdot \vec k_3$ in the numerator, but all contributions
will still have the same $N$-dependence.  This is not the
$N$-dependence of the contribution we want to cancel.  (It
is also difficult to see how one could obtain the correct $v$-dependence.)

Another possible concern is that with so many legs, it is not
straightforward to work in light cone gauge for generic momenta
and polarizations.  But by studying particular cases
one can check that this is not a problem.  For example, one can
take two of the incoming (and outgoing) polarizations
to be the same.  In this case, it is easy to see that
there are no new contributions with the same dependence
on the momenta as those which we have studied here.

It may be that there is a conceptual problem with our understanding
of the matrix model.  Perhaps there are additional contributions
which should be included which can reproduce the missing terms.
Such phenomena are familar in light-cone field theory where it
is necessary to include contact terms arising from
integrating out backward-moving particles.  In previous
calculations, such terms have been suppressed by powers
of $N$.  However, in the present case, they potentially
have the correct $N$ dependence to generate the missing
term.  We have not been able to understand in any
detail how this might work; if it is the resolution,
the question will be whether there is some
simple rule to generate these extra terms in the framework
of the matrix model.  We should note that the arguments of \cite{nati}
strongly suggest that this cannot be the source of the difficulty.
It may also be that the use of the effective Hamiltonian is problematic.
For example, the effective action presumably contains acceleration terms,
which may complicate the Hamiltonian treatment.
Finally, there is of course the
possibility that the finite $N$ matrix model, as
presently
formulated, does
not reproduce DLCQ M-theory.

One might wonder if the derivation by Dijgraff, Verlinde and
Verlinde \cite{dvv} of the three string vertex already implies that
all graviton scattering amplitudes must be reproduced correctly in the matrix
model. This is not straightforward. Light cone string theory
contains an infinite set of additional terms (the so-called
{\it contact terms}) that are not derived from the string vertex,
but instead must be put in by hand to restore Lorentz invariance
\footnote{ We are indebted to Steve Shenker for a crucial
discussion of these issues.}.
One such term
is the famous $v^4/r^7$ term that we have already referred to. The
term of interest in this paper is
another such term. The question is then whether some
property such as analyticity of the amplitude implies that
the contact terms must be reproduced correctly if the
string vertices are reproduced (along the lines of \cite{green}.) We do
not know the answer to this question.  One must also note that
higher order corrections to the three-string vertex which were
undetermined in \cite{dvv} may be
important.

The most impressive and concise argument to date for the validity
of the matrix model is that of \cite{nati}.  We do not
currently understand how this argument might be reconciled
with the apparent discrepancy we have found.

There is also an issue as to whether the discrepancy
we have found disappears in the infinite momentum limit. This
can happen because the size of the zero-brane clusters
forming a graviton grows with $N$. In the limit as $N$ tends to infinity,
the clusters are infinitely large, and there is no sense in
which the separation between the clusters can be assumed to
be large. This in turn means that we cannot find the effective
action by integrating out massive strings as we have done here.
To do the calculation in this limit, one needs more knowledge of
the bound state wave function. It is hence conceivable that the
discrepancy vanishes in the infinite momentum limit.

Note added: After the completion of this work, we received a paper by
M. Douglas and H. Ooguri \cite{do2}, which describes a discrepancy between
the matrix model and supergravity in the presence of curved
backgrounds, following the analysis of \cite{dos}. It seems likely that this
discrepancy is related to the one described in this work. 
\section{Acknowledgments}
We thank  W. Fischler, J. Polchinski, N. Seiberg,
E. Silverstein, S. Shenker,
and L. Susskind for extensive discussions.  S.
Shenker, in particular, has suggested for some
time that multigraviton scattering would provide
an interesting test of the matrix model, and W. Fischler
collaborated at an early stage.  The work
of M.D. was supported in part by the Department of
Energy, and in part by the Stanford Institute for Theoretical
Physics.  The work of A.R.
 was supported in part by the Department of Energy
under contract no. DE-AC03-76SF00515.


\end{document}